\documentclass[aps,pre,twocolumn,groupedaddress]{revtex4}
\usepackage[a4paper]{geometry}
\usepackage[utf8]{inputenc}
\usepackage{graphicx}
\usepackage{amsmath}

\graphicspath{{data/}}

\newcommand{\fig}[1]{Fig.\ \ref{#1}}

\newcommand{\eq}[1]{Eq.\ (\ref{#1})}

\newcommand{\conn}{\ensuremath{c}}

\begin{document}
\title{Distribution of diameters for Erd\H{o}s-R\'enyi random graphs}
\author{A. K. Hartmann
  \email{a.hartmann@uni-oldenburg.de}
}
\affiliation{
Institut f\"ur Physik,\\
Carl von Ossietzky Universit\"at Oldenburg,\\
26111 Oldenburg, Germany  }

\author{M. M\'ezard %
  \email{marc.mezard@ens.fr}
}\affiliation{Département de Physique de l' ENS, PSL
  Research University, 75005 Paris, France }

\date{\today}

\begin{abstract}
We study the distribution of diameters $d$ of Erd\H{o}s-R\'enyi random graphs
with average connectivity $c$. The diameter $d$ is the maximum 
among all shortest distances between pairs of nodes in a graph and
an important quantity for all dynamic processes taking place on graphs.
Here we study the distribution $P(d)$ numerically for various values of $c$,
in the non-percolating and the percolating regime. Using large-deviations
techniques, we are able to reach small probabilities like $10^{-100}$
which allow us to obtain the distribution over basically the full range
of the support,  for graphs up to $N=1000$ nodes. For values $c<1$, our
results are in good agreement with analytical results, proving the
reliability of our numerical approach.
For $c>1$ the distribution is more complex and no complete analytical results
are available. For this parameter range, $P(d)$ exhibits an inflection
point, which we found to be related to a structural change of the graphs.
For all values of $c$, we determined the finite-size rate function
$\Phi(d/N)$ 
and were able to extrapolate numerically to $N\to\infty$, indicating that
the large deviation principle holds. 
 \end{abstract}

\pacs{}

\maketitle

%
%
\section{Introduction}
\label{sec:intro}

For each connected \emph{component} $\kappa$ of 
a network or a graph $G=(V,E)$ 
\cite{cohen_book2010,newman_book2010,estrada_book2011}, 
the \emph{diameter} $d(\kappa)$ is the 
maximum, over all pairs of component's vertices $i,j$, 
of the  shortest-path distance $i \leftrightarrow j$. The diameter of
a \emph{graph}
 is the maxi\-mum over all components $\kappa$ of $d(\kappa)$.
The diameter is an important measure of the network. 
It has a strong influence on, e.g., dynamical processes 
taking place on these networks, since it characterizes a typical long 
length scale for the transport of information.
 Examples for the importance of the diameter are
rumour  spreading \cite{doerr2012}, 
energy transport in electric grids  \cite{resilience2014}
or oscillations in  neural circuits \cite{goldental2015}. Furthermore,
for networks changing over time, the temporal evolution of the
diameter can give important insights into the structure of the dynamics
\cite{blonder2012}.

Not much is known about the behaviour of the 
diameter of random network ensembles.
At least it is known that the average diameter 
for many ensembles grows logarithmically with the
number of nodes \cite{albert2002review}. 
Nevertheless, a full description, i.e.,  the probability
distribution of network diameters over the instances of an random graph 
ensemble, has almost been obtained only in very limites cases, to the knowledge
of the authors.

Thus, here we deal with the most fundamental and least-structured graph
ensemble, which are 
Erd\H{o}s-R\'enyi (ER) random graphs \cite{erdoes1960}. 
Let $N=|V|$ denote the number of vertices.
Each realisation of an ER random graph is generated by iterating
over the $N(N-1)/2$ pairs $i,j$ of nodes and adding the 
edge $\{i,j\}\in V^{(2)}$ with probability $p$. Here we concentrate
on the sparse case $p=\conn/N$, $\conn$ being the average connectivity.

In the non-percolating phase $c<1$, close to the percolation threshold 
$\conn\nearrow 1$, the distribution of
diameters is described in theorem 11(iii) of Ref.\ \cite{luczak1998}. 
The distribution is asymptotically ($N\to\infty, c\to 1$)  given by a Gumbel 
(extreme-value) distribution
\begin{equation}
P_{\rm G}(d)=\lambda e^{-\lambda (d-d_0)}e^{-e^{-\lambda(d-d_0)}}\,.
\label{eq:Gumbel}
\end{equation}
Here, $d_0$ is the maximum of the distribution, which scales
logarithmically with the number $N$ of nodes. $\lambda$ is the Gumbel parameter 
 describing the exponential behaviour $P_{\rm G}\sim e^{-\lambda d}$
for large values. It describes the variance, which is proportional to 
$1/\lambda^2$. In this $c\to 1$ limit, 
the Gumbel parameter $\lambda$
as a function of the connectivity $\conn$ is given by
\begin{equation}
\lambda(\conn) = -\log(\conn)\,.
\label{eq:prediction:lambda}
\end{equation}

The fact that $P(d)$ is connected to such an extreme-value distribution
is intuitively clear: below the percolation threshold, each graph
consists of a large number of trees, hence the diameter is obtained
by maximising over these trees.

Note that Ref. \cite{luczak1998} also
contains results for general values of $\conn<1$. Although they are
given in a more complex and partially implicit
form, they indicate that the asymptotic distribution is also the
Gumbel distribution \ref{eq:Gumbel}, with a parameter $\lambda$ also
given by \ref{eq:prediction:lambda}.

Due to finite-size corrections, the distribution of diameters in finite-size graphs does not
follow the Gumbel distribution. In earlier numerical studies 
of another problem, sequence alignments
\cite{align2002,align_long2007,newberg2008},
the data was well fitted by ``modifying'' the Gumbel distribution by a Gaussian
factor:
\begin{equation}
P_{\rm mG}(d)=\lambda'P_{\rm G}(d)e^{-a(d-d_0)^2}
\label{eq:modified:Gumbel}
\end{equation}
where $\lambda'$ is given through the normalisation 
$\int P_{\rm mG}(\delta)\,d\delta d=1$. This distribution will be used
in our analysis.

The probabilities $P(d)$ for values of $d$ which deviate 
from the typical size
are often exponentially small in $N$. Hence, one uses the concept of the
large-deviation \emph{rate function} \cite{denHollander2000,touchette2009} 
by writing

\begin{equation}
P(d) = e^{-N\Phi(d/N)+o(N)}\quad (N\to\infty)\, .
\label{eq:distr:exp:small}
\end{equation}
Note that the normalisation is part of the $e^{o(N)}$ factor.
One says that the \emph{large-deviation principle} holds if, 
loosely speaking, the empirical rate function
\begin{equation}
\Phi_N(d/N)\equiv - \frac 1 N \log P(s)
\label{eq:rate:fct}
\end{equation}
converges to $\Phi(d/N)$ for $N\to\infty$. Due to the logarithm
the normalisation and the sub-leading term of $P(s)$ become 
an additive contribution to $\Phi$, which go to zero for $N\to \infty$.

In the present work, we study numerically 
the distribution $P(d)$ of diameters of ER random graphs in the sparse regime 
$p=\conn/N$. Using a large-deviation technique which is based
on studying a biased ensemble characterised by a finite temperature-like
parameter, see Sec.\ \ref{sec:method},
we are able to obtain the distributions over almost the full 
ranges of the support,
down to very small probabilities like $10^{-100}$. For the
non-percolating regime $\conn <1$, we compare our numerical results to the available
analytical results and find a good agreement. In particular we find that
the asymptotic $\conn\nearrow 1$ result of a suitably scaled 
Gumbel distribution, modified by a Gaussian finite-size correction,
 is compatible for all values $\conn <1$ with
our results. Also, we find the dependence (\ref{eq:prediction:lambda})
of the Gumbel parameter $\lambda$ as a function of $\conn$, for $N\to\infty$.
This confirms the validity of our approach. 

We are also able to obtain $P(d)$ numerically for $\conn>1$
where no exact result is available to our knowledge. Here we find
that the distributions exhibit an inflection point. This 
leads to a first-order transition
in our finite-temperature ensemble and makes
the numerical determination of the distribution much harder.

Nevertheless, for all values of $c$, we 
determined the rate functions for various
numbers $N$ of nodes and obtained, where necessary,
 the limiting rate function via
extrapolation. In all cases we observed a good convergence, indicating
that the large-deviation principle seems to hold. 


%
%
\section{Simulation and reweighting method} 
\label{sec:method}
We are interested in determining the distribution $P(d)$ for 
the diameter of an ensemble of random graphs.  The distribution
can be obtained in principle for any graph ensemble, here we
apply it to ER random graphs.
 \emph{Simple sampling} is
straightforward: One generates a certain number $K$ of graph samples
and obtains $d(G)$ for each sample $G$.
This means each graph $G$ will appear with its natural ensemble 
probability $Q(G)$.
The probability that the graph has diameter $d$ is given by
\begin{equation}
P(d) = \sum_{G} Q(G)\delta_{d(G),d} \label{eq:PS}
\end{equation}
Therefore, by calculating a histogram of the values for $d$, an
estimation for $P(d)$ is obtained.
Nevertheless, with this simple sampling, $P(d)$ can only be measured in a regime where $P(d)$
is relatively large, about $P(d)>1/K$. Unfortunately, the distribution
usually decreases very quickly, e.g., exponentially 
in the system size $N$ when moving
away from its typical (peak) value, like in Eq.\ (\ref{eq:distr:exp:small})
This means that even for moderate system
sizes $N$, the distribution will be unaccessible through this method, on almost its complete support.

\subsection{Markov-chain Monte Carlo approach}

To estimate $P(d)$ for a much larger range of diameters,
a different \emph{importance sampling} 
approach is used \cite{align2002,largest-2011}. 
For self-containedness, the method is outlined subsequently.
The basic idea is to generate random graphs with a probability that includes an additional
Boltzmann factor $\exp(-d(G)/T)$, $T$ being a ``temperature''
parameter,  in the following manner:
A standard Markov-chain MC simulation \cite{newman1999,landau2000}
is performed, where the current state at ``time'' $t$ 
is given by an instance of a graph $G(t)$. Here the 
Metropolis-Hastings
algorithm \cite{metropolis1953} is applied  as follows: at each step
$t$ a \emph{candidate} graph $G^*$ is created from the current graph $G(t)$.
 One then computes the diameter of the candidate
graph, $d(G^*)$.
To complete a step of the Metropolis-Hastings algorithm,
the candidate graph is \emph{accepted}, ($G(t+1)=G^*$) 
with the Metropolis probability
\begin{equation}
p_{\rm Met} = \min\left\{1,e^{-[d(G^*)-d(G(t))]/T}\right\}\,.
\end{equation}
Otherwise the current graph is kept ($G(t+1)=G(t)$).  

Here, the genreation of $G^*$ is done using the following local update
rule:
A node $i\in V$ of the current graph is selected randomly, 
with uniform weight $1/N$,
and all adjacent edges are deleted. Next, the node $i$ is reconnected again:
for all other nodes $j\in V $ the
corresponding edge $\{i,j\}$
is added with a probability $c/N$ (and not added with probability $1-c/N$), 
which corresponds to its contribution to 
the natural weight $Q(G)$ of an ER graph.

By construction,
the algorithm fulfils detailed balance. Clearly the algorithm is also
ergodic, since within $N$ steps, each possible graph may be constructed. Thus,
in the limit of infinitely long Markov chains,
the distribution of graphs will follow the probability
\begin{equation}
q_T(G) = \frac{1}{Z(T)} Q(G)e^{-d(G)/T}\,, \label{eq:qT}
\end{equation}
where $Z(T)$ is the a priori unknown normalisation factor. Note that
for $T\to \infty$ all candidate graphs will be accepted and the
distribution of graphs will follow the original ER weights. 

\subsection{Obtaining the distribution}

The probability to measure $d$ at any temperature $T$ is given by
\begin{eqnarray}
P_T(d) & = &\sum_{G} q_T(G) \delta_{d(G),d} \nonumber\\
       & \stackrel{(\ref{eq:qT})}{=} & 
         \frac{1}{Z(T)}\sum_{G} Q(G)e^{-d(G)/T} \delta_{d(G),d} \nonumber \\
       & = & \frac{e^{-d/T}}{Z(T)} \sum_{G} Q(G) \delta_{d(G),d}
               \nonumber \\
       &  \stackrel{(\ref{eq:PS})}{=} & 
           \frac{e^{-d/T}}{Z(T)} P(d) \nonumber\\
\Rightarrow \quad P(d) & = & e^{d/T} Z(T) P_T(d) \, .\label{eq:rescaling}
\end{eqnarray}
Hence, the target distribution $P(d)$ can be estimated, up to a normalisation
constant $Z(T)$, from sampling at finite temperatures $T$. For each
temperature, a specific range of the distribution $P(d)$ will be sampled:
Using a positive temperature allows to sample the region of
a distribution left to its peak (values of the diameter
smaller than the typical value). Since $T$ is only an artificial
resampling parameter, also  
negative temperatures are feasible, which therefore allow
 us to access the right tail of 
$P(d)$.
In both cases, temperatures of large absolute value will cause a sampling of the 
distribution close to its typical value, while temperatures of small 
absolute value
are used to access the tails of the distribution. Hence one chooses a
suitable set of temperatures $\{T_{-N_{n}},T_{-N_{n}+1},\ldots,
T_{N_{p}-1},T_{N_{p}}\}$ with
$N_{n}$ and $N_{p}$ being the number of negative
and positive temperatures, respectively. 
A good choice of the temperatures is such that the resulting 
histograms of neighbouring temperatures overlap sufficiently. This allows
to ``glue'' the histograms together, see next paragraph.
By obtaining
the distributions $P_{T_{-N_n}}(d)$, \ldots, $P_{T_{N_p}}(d)$, such that
$P(d)$ is ``covered'' as much as possible, one can measure 
$P(d)$ over a large
range, possibly on its full support.
The range where the distribution can be obtained may be limited,
e.g., when the MC simulations at certain 
 temperatures $T_k$ do not equilibrate. This happens usually for small absolute values 
$|T_k|$, where the system might also have a glassy behaviour.
 Another difficult case is when $P(d)$ is not concave: then
a first order transition will appear \cite{largest-2011}
as a function of $T$, which might
prevent one from obtaining $P(d)$ in some regions of the support for large
systems.

The normalisation constants $Z(T)$ can easily be computed, e.g., by including
a histogram generated from simple sampling, which corresponds
to the temperature $T=\pm\infty$. 
 Using suitably chosen
temperatures $T_{+1}$, $T_{-1}$, one measures histograms which overlap 
with the simple sampling histogram on its left and right border,
respectively. Then the corresponding \emph{relative} normalisation
constants $Z_{\rm r}(T_{\pm 1})$ can be obtained by the requirement that,
after rescaling the histograms according to 
(\ref{eq:rescaling}), they must agree
in the overlapping regions with the simple sampling histogram within
error bars. This means, the histograms are ``glued'' together. In the same
manner, the range of covered 
$d$ values can be extended iteratively to the left and to
the right by choosing additional suitable temperatures 
$T_{\pm 2}, T_{\pm 3}, \ldots$ and gluing the 
resulting histograms one to the other. The histogram obtained finally
can be normalised (with constant $Z$), 
such that the probabilities sum up to one. This 
also yields the actual normalisation constants $Z(T)=Z_{\rm r}(T)/Z$ from 
Eq.\ (\ref{eq:rescaling}).
Note that one could also not only glue together neighbouring
histograms, but use for each bin value $d$
 all data which is available, as it is done, e.g.,
within the multi-histogram approach by Ferrenberg and Swendsen 
\cite{ferrenberg1989}.
For the present case, it was easy to obtain good data statistics 
such that it was sufficient to use the histograms just pairwise.

A pedagogical explanation of the gluing process
 and examples of 
this procedure can be found in Ref.\ \cite{align_book}.

\subsection{Equilibration}

In order to obtain the correct result, the MC simulations must be
equilibrated. In our case, there is an easy test that can indicate
when equilibration has not been reached. One can run two simulations starting
with two different initial graphs $G(t=0)$, respectively: 
\begin{itemize}
\item a graph which is arranged in a line, 
i.e., it has a linear structure with $N$ nodes and $N-1$ edges.
 The diameter  of this graph is $N-1$.

\item a complete graph, which contains
all $N(N-1)/2$ possible edges. For this graph
the diameter is one.
\end{itemize}

For each of these two different initial conditions,
the evolution of $d(t_{\rm MCS})$  will approach the equilibrium range 
of values from two different
extremes, which allows for a simple equilibration test:
if
the measured values of $d$ disagree within the range of fluctuation,
equilibration is not achieved. We shall assume that, conversely, if
the measured values of $d$ agree, then
equilibration has been obtained.

Note that one can also use 
generalised ensemble methods like the Multicanonical method \cite{berg1992}
or the  Wang-Landau approach \cite{wang2001}, in particular when a
first order transition as function of $T$ appears,
 to obtain the distribution $P(d)$. While these methods in principle
 do not require to perform independent
simulations at different values for the temperatures,
it turns out that for larger system sizes, one still has to perform multiple
simulations because one has to split the interval of interest into
smaller overlapping sub intervals, to make the simulation feasible.
Here, the Wang-Landau algorithm was used only for the case of $c=3$,
where the temperature-based approach did not work well, see below.
For other values of $c$, it turned out to be much easier to guide the simulations
to the regions of interest, e.g., where data is missing
using the so-far-obtained data, and to monitor the equilibration process, using the finite-temperature
approach.

%
%
\section{Results}
\label{sec:results}

We have numerically determined the distribution of diameters for
ER random graphs for different connectivities below, above and at
the percolation threshold $c_c=1$.

\subsection{Connectivity $c<1$}

\begin{figure}[htb]
  \centering
  \includegraphics[clip,width=0.45\textwidth]{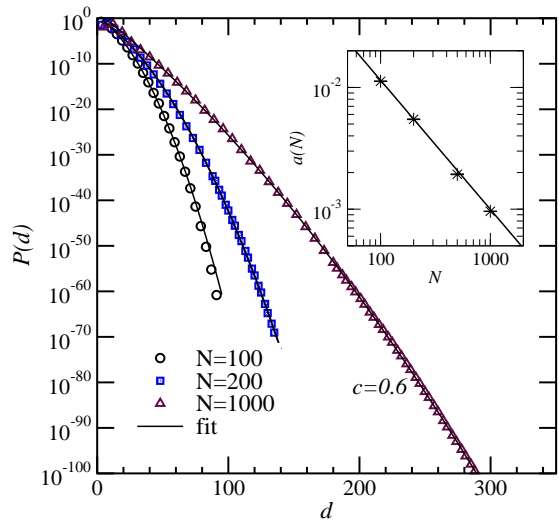}
  \caption{
    \label{fig:Pd_er06}
     Distribution of diameters for Erd\H{o}s-R\'enyi random graphs with
average connectivity $\conn=0.6$ for three different graphs sizes $N=100$,
$200$, and $1000$. The lines show fits to the ``modified'' Gumbel distribution
according to \eq{eq:modified:Gumbel}. The inset shows the dependence
of the parameter $a$ of \eq{eq:modified:Gumbel} on the system size $N$.
The line displays the power law $\sim 1.51 N^{-1.06}$.}
\end{figure}

We start with the non-percolating regime, where we can compare with
exact asymptotic results
\cite{luczak1998}. In Fig.\ \ref{fig:Pd_er06}, $P(d)$ is shown
at $c=0.6$ for three different graph sizes. Using the approach
explained in the last section,  probabilities as small as
$100^{-100}$ are easily accessible. In the log-log plot, clearly a curvature in\
the data is visible for large values of the diameter, which could be
partially due to
strong finite-size corrections.
 We have fitted the data to a modified Gumbel distribution
\eq{eq:modified:Gumbel} and obtained good fit qualities.
We studied the strength $a$ of the Gaussian correction, see
inset of Fig.\ \ref{fig:Pd_er06}. One observes a clear power-law behaviour.
 Hence, the numerical data supports that
asymptotically the full distribution becomes Gumbel or Gumbel-like 
below the percolation threshold.

\begin{figure}[htb]
  \centering
  \includegraphics[clip,width=0.45\textwidth]{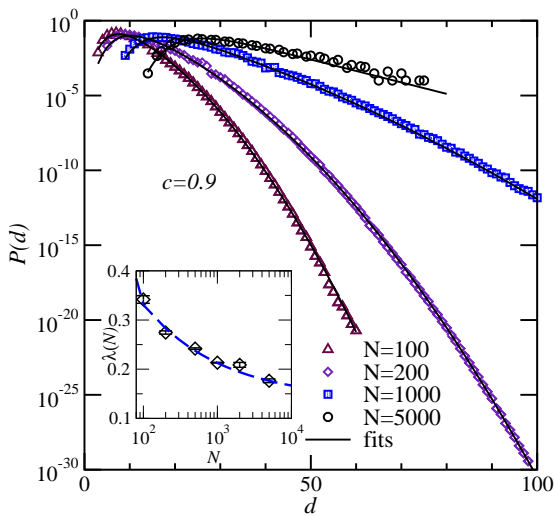}
  \caption{
    \label{fig:Pd_er09}
     Distribution of diameters for Erd\H{o}s-R\'enyi random graphs with
average connectivity $\conn=0.9$ for three different graphs sizes $N=100$,
$200$, and $1000$. The lines show fits to the ``modified'' Gumbel distribution
according to \eq{eq:modified:Gumbel}. The inset displays the dependence
of the Gumbel parameter $\lambda$ as a function of graph size $N$. The inset
shows the result of a fit to the function \eq{eq:extrapolation:lambda}.}
\end{figure}

We have studied the behaviour in the non-percolating phase for various
values of the connectivity $\conn$ and different system sizes,
see Fig.\ \ref{fig:Pd_er09} for $\conn=0.9$. Each time
we observe qualitatively the same convergence to a Gumbel
distribution. 

From the fits, for each value of the connectivity $c$ and each
system size $N$, a value of $\lambda(\conn,N)$ is obtained.
To extrapolate the dependence of the Gumbel parameter $\lambda(\conn)$ to large
graph sizes, the following heuristic dependence, inspired by standard
finite-size scaling \cite{cardy1988,goldenfeld1992} was applied:
\begin{equation}
\lambda(\conn,N)=\lambda(c)+aN^{-\alpha}\,.
\label{eq:extrapolation:lambda}
\end{equation}

\begin{figure}[hbt]
  \centering
  \includegraphics[clip,width=0.45\textwidth]{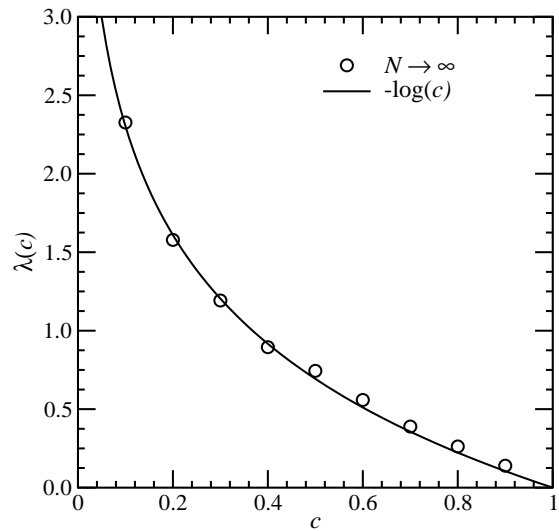}
  \caption{
    \label{fig:lambda_c_er}
    Dependence of the Gumbel parameter $\lambda$ as a function of the
connectivity $\conn$. The symbols show the extrapolation of the
numerical results using \eq{eq:extrapolation:lambda}. The error bars
are of order of symbol size.
The solid line represents
the mathematical result \eq{eq:prediction:lambda} of Ref.\ \cite{luczak1998}.  }
\end{figure}

The inset of Fig.\ \ref{fig:Pd_er09} shows the behaviour of 
$\lambda(\conn=0.9,N)$
together with the fit as function of graph size $N$. The resulting
values for $\lambda$ as a function of the connectivity $\conn$ are
shown in Fig.\ \ref{fig:lambda_c_er} together with the
asymptotic result \eq{eq:prediction:lambda}, yielding a 
nice agreement. This shows that indeed the numerical approach
allows to reliably study the distribution of diameters for finite sizes
and to extrapolate to large graphs.

\begin{figure}[htb]
  \centering
  \includegraphics[clip,width=0.45\textwidth]{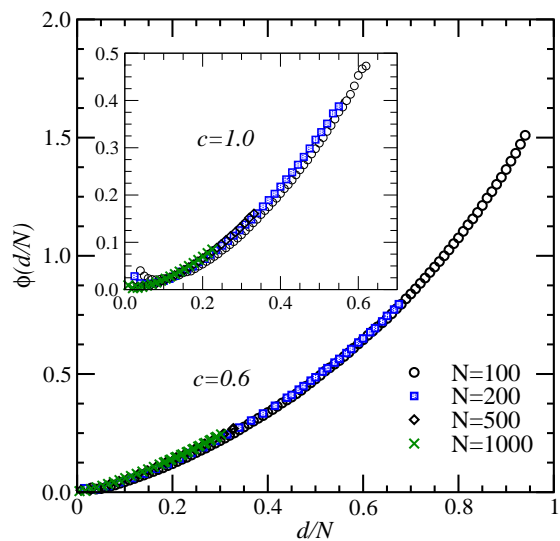}
  \caption{
    \label{fig:rate_er0.6_1.0}
Large-deviation rate function $\Phi$ as a function of the
rescaled diameter $d/N$ for $c=0.6$ and different system sizes.
The inset shows the  large-deviation rate function 
for the percolation threshold $c=1$. }
\end{figure}

Nevertheless, the scaling of the Gaussian correction parameter is basically
close to $a\sim 1/N$, hence when looking at the data 
for the rescaled diameter $\hat d =d/N$, the size-dependence
exactly drops out. Hence, the
 rate function \eq{eq:rate:fct}
as a function of $\hat d$ is studied next, 
as displayed in \fig{fig:rate_er0.6_1.0}.  
The data collapse is good, which means that even for small system sizes
the rate function is well converged.
This means that the distribution of diameters can indeed be described well
by a rate function, hence the large-deviation principle 
\cite{dembo2010,denHollander2000} holds. This makes it a bit more likely
that this model is accessible to the mathematical tool of the large-deviation
theory.
Note that due to the curvature of the rate function, the Gumbel
distribution is not visible when inspecting the result on
the $d/N$ scale, which makes the most-important finite-size
contribution drop out.  Only when one looks at the data at fixed
values $d$, the convergence to a Gumbel is a meaningful statement.
A similar result has been observed previously for the distribution
of scores of sequence alignment \cite{newberg2008}.

\subsection{Connectivity $c=1$}

\begin{figure}[htb]
  \centering
  \includegraphics[clip,width=0.45\textwidth]{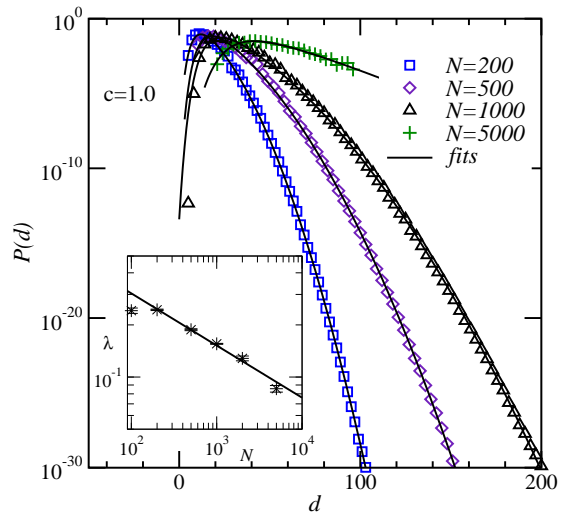}
  \caption{
    \label{fig:Pd_er10}
     Distribution of diameters for Erd\H{o}s-R\'enyi random graphs with
average connectivity $\conn=1.0$ for different graph sizes. The lines show fits 
according to \eq{eq:modified:Gumbel}. The inset shows the dependence
of the fitting parameter $\lambda$ as a function of graph size using
a double-logarithmic axis. The lines shows the power-law $\sim N^{-\alpha}$
with $\alpha=-0.3$ obtained from fitting the $\lambda(N)$ data for $N\ge 200$.
}
\end{figure}

The results for the non-percolating phase
 gave us confidence that the numerical method
allows us to turn our attention to cases,
where no exact results for the full distribution are available.
In Fig.\ \ref{fig:Pd_er10}, the distribution of diameters is shown
right at the percolation transition $\conn=1$. Here, 
the random graph consists of a large extensive
tree plus small components and no Gumbel distribution is expected,
since $\lambda=0$.

Nevertheless, it is still possible to fit the finite-size data
to the modified Gumbel distribution, see Fig.\ \ref{fig:Pd_er10},
since any finite-size graph for $\conn=1$ cannot
be distinguished from the case $\conn$ close to 1.
 For example, fitting the case $N=1000$ to \eq{eq:modified:Gumbel}
resulted in
\begin{equation*}
\lambda=0.151(1), d_0=22.9(1), 
a=1.304(6)\times 10^{-3}\,.
\end{equation*}
The fit matches the data well, also for other systems sizes. Nevertheless,
when studying the dependence of $\lambda$ on the system size,
a convergence towards zero seems most likely, see inset of Fig.\ 
\ref{fig:Pd_er10}. In the double-logarithmic plot
the data appears to be compatible with a straight line, meaning a power-law
decrease, and maybe even a faster decrease. We verified this
by fitting $\lambda(N)$ according to
\eq{eq:extrapolation:lambda}, where we obtained a negative value
for $\lambda$ for $N\to\infty$ with an error bar of almost the
same size, showing that indeed
the distribution $P(d)$ differs from the Gumbel distribution
for $\conn=1$.

This can also be seen from studying the large-deviation rate function
$\Phi(d/N)$,
see inset of \fig{fig:rate_er0.6_1.0} where also an upward-bending function
is seen, as for the case $c=0.6$.


\subsection{Connectivity $c>1$}

\begin{figure}[htb]
  \centering
  \includegraphics[clip,width=0.45\textwidth]{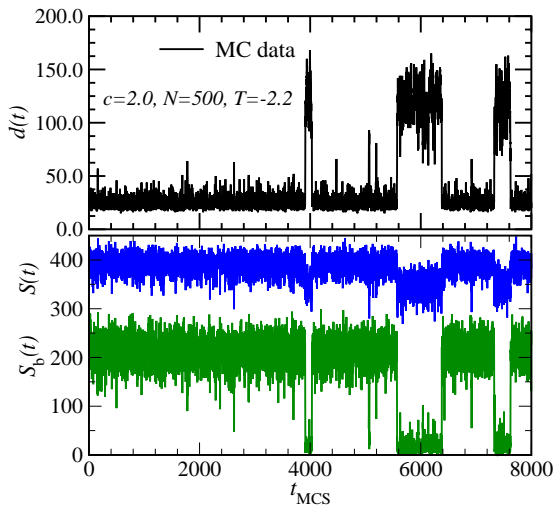}
  \caption{
    \label{fig:d:t}
Top: Time series for the diameter $d$ as a function of the number
$t_{\rm MCS}$ of MC sweeps, for $\conn=2$, graph size $N=500$
at artificial temperature $T=-2.2$. Bottom: for the same run,
the size $S$ of the largest component and the size $S_{\rm b}$ of the largest
biconnected component as a function of $t_{\rm MCS}$.}
\end{figure}

In the percolating phase $\conn>1$, the numerical
result show that in the artificial finite-temperature
ensemble there appears to be a 1st order
phase transition as a function of the temperature, similarily to
the distribution of the size of the largest
component \cite{largest-2011}. 
To visualise this, we here only show an example
of a MC time series for the diameter, see Fig. \ref{fig:d:t}. Clearly
the diameter oscillates between two distinct regimes, showing the
coexistence of two ``phases' with small and large diameter, respectively.
This corresponds at this temperature 
to a distribution of diameters $P_T(d)$ with two peaks.

The data in Fig.\ \ref{fig:d:t} 
of the size $S$ of the largest component and the size $S_{\rm b}$ 
of the largest biconnected component 
suggest that the system oscillates between two states. 
When the diameter is small, around 30 here, the largest biconnected component
is large, it contains about 200 nodes. 
On the other hand, when the diameter is large, about 130, the largest 
biconnected component has a size of only about 30. Nevertheless, the size of
the largest components changes only a little bit. This we interpret in the 
following way.

\begin{figure}[t!]
  \centering
  \includegraphics[clip,width=0.45\textwidth]{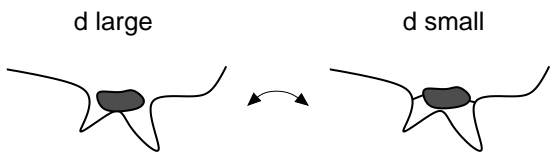}
  \caption{
    \label{fig:two_structures}
The bimodal characteristics of the graphs at phase coexistence ($c>1$):
There is a large linear structure plus a tightly connected structure
(plus a lot of small components which are not so important).
At criticality, the system oscillates between states where the 
strongly connected structure is attached several times 
to the linear structure, allowing for shortcuts, or only weakly connected
(sometimes even not at all).
}
\end{figure}

There
is always one large line-like object present and a tightly (bi-) 
connected 
cluster,  see Fig.\ \ref{fig:two_structures}.
As we have verified explicitly in our numerical data, 
the tightly connected cluster typically has a small diameter while
the line-like object has a large diameter.
In one state, the line-like object is connected
to the tightly connected cluster only at few nodes or not at all.
Hence, the diameter path is basically along the line-like object
and the diameter is large.
In the other state, the line-like object is connected to the tightly-connected
cluster at several, distant points, such that the diameter path
makes a shortcut. Thus, the diameter is small and the biconnected component
relatively large.
Our measurements showed that although the diameters of the two states 
differ strongly, the number of edges 
differ only slightly (not shown here).
This allows for a quick transition between
the two states.  Note that in the mathematical literature 
\cite{chung2001}
for a logarithmically growing connectivity for diluted 
ER random graphs, the  diameter is
concentrated around a finite number (larger than one) of values. This might be
related to the observed oscillations of the diameter.

The existence of an inflection points translates into 
the existence of a first-order transition in the finite
temperature ensemble with some criticial temperature $T_{\rm c}$.
This leads to a bimodal structure of $P_{T_c}(d)$ exhibiting a very small
probability in the region between the two peaks.
Hence, concerning the numerical effort,
obtaining the full distribution $P(d)$ becomes difficult, in
particular for large graphs, because the intermediate region for
values of $d$ between the two peaks of $P_T(d)$ 
is rarely or even not at all sampled.

\begin{figure}[htb]
  \centering
  \includegraphics[clip,width=0.45\textwidth]{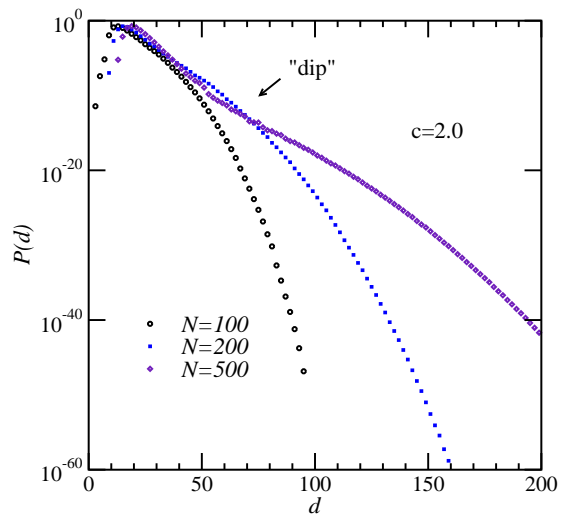}
  \caption{
    \label{fig:Pd_er20}
     Distribution of diameters for Erd\H{o}s-R\'enyi random graphs with
average connectivity $\conn=2.0$ for graph size 
$N=100$, $200$ and $500$. }
\end{figure}

Therefore, for the case $\conn=2.0$, only system sizes up to $N=500$ could
be equilibrated deep into the large-diameter regime (corresponding
to negative temperatures with small absolute value). The first-order
nature of the transition, i.e., the two-peak structure of the
distributions of the diameter at finite-temperature close
to the transition temperature, becomes visible by a ``dip'' in the distribution
of diameters for $N=500$.

\begin{figure}[htb]
  \centering
  \includegraphics[clip,width=0.45\textwidth]{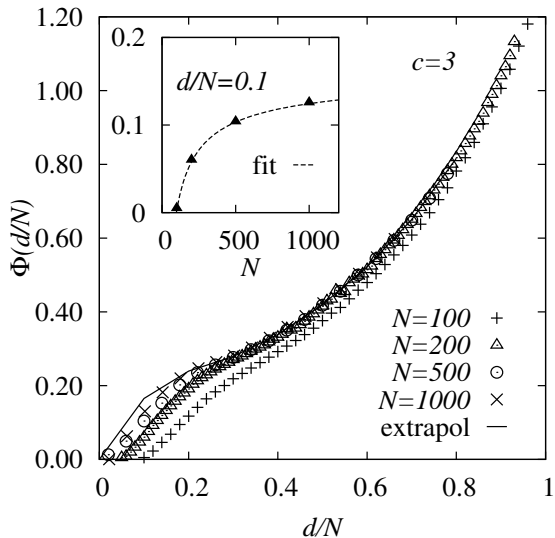}
  \caption{
    \label{fig:rate_er3.0}
Large-deviation rate function $\Phi$ as a function of the
rescaled diameter $d/N$ for $c=3.0$ and different system sizes
$N=100,200,500, 1000$ (symbols). The line shows the rate function obtained
from extrapolation $N\to\infty$.
Inset: extrapolation of rate function as function
of system size, for $d/N=0.1$.
 }
\end{figure}

The dip is more pronounced when going to larger connectivities.
This can be seen in \fig{fig:rate_er3.0}, where the large-deviation rate
function is displayed for $\conn=3.0$ 
for different system sizes. Here, the Wang-Landau 
approach was used, which allowed us to sample the region of the
dip much better.

In Fig.\ \ref{fig:rate_er3.0} the corresponding rate function is shown.
Here, in particular for small values of $d/N$, stronger finite size effects
are visible. Thus, we have considered the functions $\Phi_N(d/N)$
at various fixed rations $\tilde d = d/N$ and performed an extrapolation via
fitting
\begin{equation}
\Phi_N(\tilde d)=\Phi(\tilde d)+bN^{-\beta}\,,
\end{equation}
where $b$ and $\beta$ are fitting parameters which are determined for each
considered value of $\tilde d$ separately, i.e., point-wise. 
An example of the extrapolation
is shown in Fig.\ \ref{fig:rate_er3.0}, together with the extrapolated values $\Phi(d/N)$.
For large values of $d/N$, above 0.5, the
small finite size effects are small, while for small values of $d/N$ the extrapolated
function differs slightly from the results for finite values of $N$.
The change from a concave to a convex function near $d/N=0.3$ is well visible.
A similar qualitative behaviour has been found previously for the rate function $\Phi$
for the distribution of the size of the largest component
for ER random graphs.

\section{Summary}

We have studied the distribution of the diameter for dilute ER random graphs
with connectivities $c$.
Using large-deviations simulations techniques, we were able to obtain
the distributions over large ranges of the support. In the non-percolating 
region of small connectivities $c<1$, the distributions are concave and
can be well fitted to the Gumbel distributions with a Gaussian correction.
The extrapolated parameter of the Gumbel distribution agrees well with
mathematical results. In the percolating regime $c>1$ the distribution
of the diameters is not available. Within the numerical result, we
observed a change from concave to convex behaviour, thus a more complex
distribution. Nevertheless, for all values of $c$ we studied, we were
able to obtain and extrapolate the rate function. This means that the
distribution of diameters follows the large deviation principle.

Since the diameter is of importance for many physical processes
taking place on networks, it would be interesting to obtain the
distribution over a large range of the support for other graph ensembles,
like scale free graphs. The results obtained in the present work show
that this should  in principle be possible.

\section*{Acknowledgements}
We thank Charlotte J. Beelen for many valuable discussions and a critical
reading the manuscript.
 This project was  supported  by  the  German {\em  Humboldt
   foundation}.
MM thanks the University of Oldenburg for the hospitality.
AKH is grateful for the hospitality and the financial support of the
LPTMPS, Universit\'e Paris-Sud, were he spent part of his sabbatical.
The simulations were performed on the HERO cluster of the University
of Oldenburg jointly funded by the DFG (INST 184/108-1 FUGG) and the 
ministry of Science and Culture (MWK) of the Lower Saxony State.

\bibliography{diameter}

\end{document}